\documentclass[a4paper,10pt]{article}
\usepackage{mathtext}
\usepackage[T2A]{fontenc}
\usepackage{amsfonts}
\usepackage{amsmath}
\usepackage{amsthm}
\usepackage{amssymb}
\usepackage{multirow}

\renewcommand{\arraystretch}{1.3}
\makeatletter
\newdimen\normalarrayskip              
\newdimen\minarrayskip                 
\normalarrayskip\baselineskip \minarrayskip\jot
\newif\ifold             \oldtrue            \def\new{\oldfalse}
\def\arraymode{\ifold\relax\else\displaystyle\fi} 
\def\eqnumphantom{\phantom{(\theequation)}}     
\def\@arrayskip{\ifold\baselineskip\z@\lineskip\z@
     \else
     \baselineskip\minarrayskip\lineskip2\minarrayskip\fi}
\def\@arrayclassz{\ifcase \@lastchclass \@acolampacol \or
\@ampacol \or \or \or \@addamp \or
   \@acolampacol \or \@firstampfalse \@acol \fi
\edef\@preamble{\@preamble
  \ifcase \@chnum
     \hfil$\relax\arraymode\@sharp$\hfil
     \or $\relax\arraymode\@sharp$\hfil
     \or \hfil$\relax\arraymode\@sharp$\fi}}
\def\@array[#1]#2{\setbox\@arstrutbox=\hbox{\vrule
     height\arraystretch \ht\strutbox
     depth\arraystretch \dp\strutbox
     width\z@}\@mkpream{#2}\edef\@preamble{\halign
\noexpand\@halignto
\bgroup \tabskip\z@ \@arstrut \@preamble \tabskip\z@ \cr}%
\let\@startpbox\@@startpbox \let\@endpbox\@@endpbox
  \if #1t\vtop \else \if#1b\vbox \else \vcenter \fi\fi
  \bgroup \let\par\relax
  \let\@sharp##\let\protect\relax
  \@arrayskip\@preamble}
\def\eqnarray{\stepcounter{equation}%
              \let\@currentlabel=\theequation
              \global\@eqnswtrue
              \global\@eqcnt\z@
              \tabskip\@centering
              \let\\=\@eqncr
 \halign to \displaywidth\bgroup
    \eqnumphantom\@eqnsel\hskip\@centering
    $\displaystyle \tabskip\z@ {##}$%
    \global\@eqcnt\@ne \hskip 2\arraycolsep
         $\displaystyle\arraymode{##}$\hfil
    \global\@eqcnt\tw@ \hskip 2\arraycolsep
         $\displaystyle\tabskip\z@{##}$\hfil
         \tabskip\@centering
    &{##}\tabskip\z@\cr}
\begingroup\ifx\undefined\newsymbol \else\def\input#1 {\endgroup}\fi

\catcode`\@=11
\def\marginnote#1{}

\newcount\hour
\newcount\minute
\newtoks\amorpm
\hour=\time\divide\hour by60 \minute=\time{\multiply\hour by60
\global\advance\minute by-\hour}
\edef\standardtime{{\ifnum\hour<12 \global\amorpm={am}%
        \else\global\amorpm={pm}\advance\hour by-12 \fi
        \ifnum\hour=0 \hour=12 \fi
        \number\hour:\ifnum\minute<10 0\fi\number\minute\the\amorpm}}
\edef\militarytime{\number\hour:\ifnum\minute<10 0\fi\number\minute}

\def\draftlabel#1{{\@bsphack\if@filesw {\let\thepage\relax
      \xdef\@gtempa{\write\@auxout{\string
          \newlabel{#1}{{\@currentlabel}{\thepage}}}}}\@gtempa \if@nobreak
    \ifvmode\nobreak\fi\fi\fi\@esphack} \gdef\@eqnlabel{#1}}
    \def\@eqnlabel{}
\def\@vacuum{}
\def\draftmarginnote#1{\marginpar{\raggedright\scriptsize\tt#1}}

\def\draft{
  \oddsidemargin -.5truein
  \def\@oddfoot{\footnotesize \sl preliminary draft \hfil
    \rm\thepage\hfil\sl\today\quad\militarytime}
  \let\@evenfoot\@oddfoot \overfullrule 3pt
    \let\label=\draftlabel
    \let\marginnote=\draftmarginnote
  \def\@eqnnum{(\theequation)\rlap{\kern\marginparsep\tt\@eqnlabel}%
    \global\let\@eqnlabel\@vacuum}

  }

\def\be{\begin{eqnarray}}
\def\ee{\end{eqnarray}}

\def\beq{\begin{equation}}
\def\eeq{\end{equation}}
\def\ba{\beq\new\begin{array}{c}}
\def\ea{\end{array}\eeq}
\def\be{\ba}
\def\ee{\ea}

\newcommand{\eq}[2]{\begin{equation} \label{#1} #2 \end{equation}}

\newcommand{\br}[1]{\left ( #1 \right )}
\newcommand{\bv}{\textbf{v}}
\newcommand{\ok}{\overline{k}}
\title{On coincidence of Alday-Maldacena-regularized $\sigma$-model and Nambu-Goto areas of minimal surfaces}
\author{{\bf A. Popolitov}\footnote{E-mail: \ popolitov@itep.ru}
\date{ } \\
{\small {\it ITEP, Moscow, Russia}}\\ \\
}
\begin{document}

\maketitle

\vspace{-7 cm}

\begin{center}
\hfill ITEP/TH-38/07\\
\end{center}

\vspace{5.0cm}

\begin{abstract}
 For the $\sigma$-model and Nambu-Goto actions, values of the Alday-Maldacena-regularized actions are calculated on solutions of the equations of motion with constant non-regularized Lagrangian. It turns out that these values coincide up to a factor, independent of boundary conditions.
\end{abstract}

\section{Introduction}
\par The idea of string/gauge theory duality is far from being new (see \cite{Schrom}, \cite{Baez} for intoduction to the subject). A particular incarnation of such a duality, namely the AdS/CFT correspondence, first stated in \cite{Mald1} and then reformulated in \cite{GubKlebPol} and \cite{Witt1}, proved to be very helpful in obtaining results in strongly coupled gauge theories\footnote{
	For breef historical review on the AdS/CFT correspondence, see \cite{Kleb1}. Quite intuitition-developing view on the topic can be found in \cite{Mor1}.
}.
\par
  Alday and Maldacena \cite{AM} have recently calculated the four-point gluon scattering amplitude in $\mathcal{N} = 4$ SYM. The point is, that at the string theory side of the duality at the zeroth order of semiclassical expansion the scattering amplitude is equal to $e^{i S_{cl}}$. Here, $S_{cl}$ stands for the action of a string, evaluated on some solution of the classical equations of motion with certain boundary conditions. Alday and Maldacena chose the $\sigma$-model action, and found some solution to the equations of motion. However, they realized that the Lagrangian is constant on this solution, so that $S_{cl}$ diverges and seems to be independent of boundary conditions. They proposed a regularization procedure in order to make answers finite. After regularization was applied, the dependence on boundary conditions was recovered.

\par
Later, Mironov, Morozov and Tomaras in \cite{MMT} and \cite{MMT2} found a whole family of solutions with a constant Lagrangian for the $\sigma$-model and Nambu-Goto actions, which are of the form $z = \Sigma_a z_a e^{\overline{k}_a u}, \textbf{v} = \Sigma_a \textbf{v}_a e^{\overline{k}_a u}$, where certain conditions are imposed on $z_a, \textbf{v}_a\ $and $\overline{k}_a$ (here, $z$ and $\textbf{v}$ are convinient coordinates on the $AdS_5$ space, $z_a, \textbf{v}_a\ \and \overline{k}_a$ are, respectively, constant scalars, 4-vectors and 2-vectors). It is known that solutions of the non-regularized NG equations of motion are solutions of the non-regularized $\sigma$-model equations of motion too, and these solutions correspond to Alday-Maldacena's choise of $z_a$. It seems that the regularization can break this simple relation between these two actions, hence the regualrized areas, a priori, can be different. In this paper, we calculate these areas explicitly and it turns out that they coincide up to a factor, dependent only on regularization parameter $\epsilon$ and angle $\phi$ between $\overline{k}_a$. This is quite surprising and seemingly indicates presense of a more subtle relationship between the $\sigma$-model and NG actions.

\par Throughout this paper we freely use the notations and results of  \cite{MMT} and \cite{MMT2}.

\section{$\sigma$-model case}
  The action under consideration is of the form
\eq{smodelaction}{
	S_\sigma = \int  G_{ij} \delta^{ij} d^2 u,
}
where $G_{ij} = \br{ \partial_i z \partial_j z + (z \partial_i \bv - \bv \partial_i z, z \partial_j \bv - \bv \partial_j z)}z^2$ - the metric induced by mapping $\mathbb{R}^2 \rightarrow AdS_5$. Following the regularization procedure given in \cite{AM}, we substitute $z \rightarrow z \br {1 + \frac{\epsilon}{2}}^{-1/2}$ and add a factor of $z^\epsilon$. Further, we use the fact that $G_{ij} = const$ on solutions, in order to exclude $\bv$
\eq{S_sigma_reg}{
	S_{\sigma \epsilon} = \br{1 + \frac{\epsilon}{2}}^{-1-\epsilon /2} \int \br{tr G + \frac{\epsilon}{2} tr \br {\frac{\partial_i z \partial_j z}{z^2}}} z^\epsilon d^2 u
}  
As soon as we are interested only in terms which do not go to zero while $\epsilon \rightarrow 0$ , in what follows all the equalities should be understood as right up to terms infinitesimal in the limit of $\epsilon \rightarrow 0$ . Solutions of the undeformed equations of motion are of the form: $z = \mathop{\Sigma}_{a} z_a e^{\ok_a u}, \bv = \mathop{\Sigma}_{a} \bv_a e^{\ok_a u}$, where $\ok^2 = tr G$. The integral $\int z^\epsilon d^2 u$ was evaluated in \cite{MMT} and is equal to
\eq{z^epsilon}{
	\int z^\epsilon d^2 u = \frac{1}{|[\ok_1 , \ok_2]|} \br { \frac{4}{\epsilon^2} + \frac{1}{\epsilon} \ln (z_1 z_2 z_3 z_4) - \frac{\pi^2}{3} + \frac{1}{8} \br {\ln^2 \br {z_1 z_2 z_3 z_4} - \ln^2 \br { \frac{z_1 z_3}{z_2 z_4}}}}
}
Thus, one immediately obtains
\be \int tr \br {\frac{\partial_i z \partial_j z}{z^2}} z^\epsilon d^2 u = \frac{1}{\epsilon (\epsilon - 1)}\mathop{\Sigma}_{a,b,i} k_i^a k_i^b z_a z_b \frac{\partial^2}{\partial z^a \partial z^b} \int z^ \epsilon d^2u =  \\
= \frac{1}{\epsilon (\epsilon - 1)}\mathop{\Sigma}_{a,b}\frac{(\ok^a, \ok^b)}{|[\ok^1, \ok^2]|} z_a z_b \frac{\partial^2}{\partial z^a \partial z^b} \int z^ \epsilon d^2u \bigg |_{\ok^2 = 1, (\ok^1, \ok^2) = 0} \label{tr z^epsilon} \ee 
Writing down the matrices $(\ok_a, \ok_b)$ and $z_a z_b \frac{\partial^2}{\partial z^a \partial z^b} \int z^ \epsilon d^2u \bigg |_{\ok^2 = 1, (\ok^1, \ok^2) = 0}$ it is easy to see that
 \be (\ref{tr z^epsilon}) = \frac{4}{|\sin (\phi)|\br{1 - \epsilon} \epsilon^2} \br {1 + \frac{\epsilon}{4} \ln (z_1 z_2 z_3 z_4)} \label {tr z^epsilon final} \ee
Since $\br{1 + \frac{\epsilon}{2}}^{-1 - \epsilon / 2} = 1 - \frac{\epsilon}{2},$ , substituting (\ref{tr z^epsilon final}) and (\ref{z^epsilon})  into (\ref{S_sigma_reg}) gives
\be S_{\sigma \epsilon} = \frac{\br {1 - \frac{\epsilon}{2}}}{|\sin (\phi)|} \br { \frac{4}{\epsilon^2} + \frac{1}{\epsilon} \ln \br{ z_1 z_2 z_3 z_4} - \frac{\pi^2}{3} + \frac{1}{8} \br{\ln^2\br{z_1 z_2 z_3 z_4} - \ln^2 \br{\frac{z_1 z_3}{z_2 z_4}}} + \frac{2}{\epsilon} + \frac{1}{2} \ln \br {z_1 z_2 z_3 z_4} + 2} = \\
= \frac{4}{|\sin (\phi)| \epsilon^2} \br {1 + \epsilon^2 \br {\frac{1}{4} - \frac{\pi^2}{12}}} \br { 1 + \frac{\epsilon}{4} \ln \br {z_1 z_2 z_3 z_4} + \frac{\epsilon^2}{32} \br { \ln^2 \br {z_1 z_2 z_3 z_4} - \ln^2
	\br{\frac{z_1 z_3}{z_2 z_4}}}} = \label{S_sigma_epsilon_final} \\
=\frac{2}{|\sin (\phi)| \epsilon^2}\br {1 + \epsilon^2 \br {\frac{1}{4} - \frac{\pi^2}{12}}} \br { (z_1 z_3)^{\epsilon/2} + (z_2 z_4)^{\epsilon / 2} - \frac{\epsilon^2}{8} \ln^2 \br {\frac{z_1 z_3}{z_2 z_4}}} \ee

Following \cite{AM}, to get a regularized area, this expression should be multiplied by
\eq{AM_factor}{
	\frac{\sqrt{\lambda_D c_D}}{2 \pi a^\epsilon} = \frac{\sqrt{\lambda \mu^{2 \epsilon}} (2 \pi)^\epsilon}{2 \pi a^\epsilon} \sqrt{1 + \epsilon + \frac{\pi^2 \epsilon^2}{12}}
}
Thus, the regularized area for the $\sigma$-model is equal to\footnote{We still reproduce the original Alday-Maldacena result, cf. \cite{NacSh}}
\eq{Area_epsilon_sigma}{
	Area_{\sigma} = \frac{1}{ \pi \epsilon^2 |\sin (\phi)|}\br{ 1 + \frac{\epsilon}{2} \br { 1 - \ln 2} + \frac{\epsilon^2}{8} \br{1 - \frac{\pi^2}{3} - 2 \ln 2 + \ln^2 2}} \br { \sqrt { \frac{ \lambda \mu^{2 \epsilon}}{(-s)^\epsilon}} + 
	\sqrt { \frac{ \lambda \mu^{2 \epsilon}}{(-t)^\epsilon}} - \frac{ \epsilon^2 \sqrt{\lambda}}{8} \ln^2 \br{\frac{s}{t}}}
}

\section{Nambu-Goto case}
In this case the action has the form
\eq{S_NG}{
	S_{NG} = \int \sqrt { \frac{1}{2} \varepsilon^{ik} \varepsilon^{jl} G_{ij} G_{kl}} d^2 u
}
Solutions with constant $G_{ij}$ look very similar to those of  $\sigma$ - model: they have the same form, but different constraints are imposed on $z_a, \bv_a, \ok_a$ (see \cite{MMT2}). Expanding in powers of epsilon, one obtains for the regularized action

\eq{S_NG1}{
	S_{NG} = \br{1 + \frac{\epsilon}{2}}^{-1-\epsilon /2} \int \sqrt{det G} \br {1 + \frac{\epsilon}{4} G^{-1 kl} \frac{\partial_k z \partial_l z}{z^2} - \frac{\epsilon^2}{32} G^{-1 kl} G^{-1 ij} \frac{ \partial_k z \partial_l z \partial_i z \partial_j z}{z^4}} z^ \epsilon d^2 u
}
It is easy to see that the action is invariant under coordinate transformations of the worldsheet $u^{'} = f (u)$. Hence, without loss of generality, we put $\ok_1 = (1, 0), \ok_2 = (0, 1), G_{ij} = diag \br{\frac{1}{2}, \frac{1}{2}}$ 
\eq{S_NG2}{
	 (\ref{S_NG1}) = \frac{1}{2} \br{1 + \frac{\epsilon}{2}}^{-1-\epsilon /2} \int \br {1 + \frac{\epsilon}{2} tr \br{\frac{\partial_i z \partial_j z}{z^2}} - \frac{\epsilon^2}{8} tr^2 \br {\frac{\partial_i z \partial_j z}{z^2}}} z^\epsilon d^2 u
}
Difference with the $\sigma$ - model case is in overall factor of $\frac{1}{2}$ and in the presense of a term contributing to the finite (i.e. $\epsilon$-independent) part of the action. This term is equal to
\eq{tr^2}{
	\int tr^2 \br {\frac{\partial_i z \partial_j z}{z^2}} z^\epsilon d^2 u = - \frac{4 \br {1  - \frac{ \pi^2 \epsilon^2}{12}}}{\epsilon (1 - \epsilon)(2 - \epsilon)(3 - \epsilon) \epsilon^2} \frac{ \epsilon}{4} \mathop{\Sigma}_{a,b} z_a^2 z_b^2\frac{ \partial^4}{ \partial z_a^2 \partial z_b^2} \ln (z_1 z_2 z_3 z_4) = \frac{4}{\epsilon^2}
}
Substituting (\ref{tr^2}), (\ref{tr z^epsilon final}) and (\ref{z^epsilon}) into (\ref{S_NG2}) and manipulating with the expression just as in the $\sigma$-model case, one obtains
\eq{S_NG3}{
	S_{NG} = \frac{1}{\epsilon^2} \br {1 + \epsilon^2 \br {\frac{1}{8} - \frac{\pi^2}{12}}}\br { (z_1 z_3)^{\epsilon/2} + (z_2 z_4)^{\epsilon / 2} - \frac{\epsilon^2}{8} \ln^2 \br {\frac{z_1 z_3}{z_2 z_4}}}
}
Thus, the answer for the regularized area in the NG case is
\eq{Area_NG}{
	Area_{NG} = \frac{1}{2 \pi \epsilon^2}\br{ 1 + \frac{\epsilon}{2} \br { 1 - \ln 2} + \frac{\epsilon^2}{8} \br{ - \frac{\pi^2}{3} - 2 \ln 2 + \ln^2 2}} \br { \sqrt { \frac{ \lambda \mu^{2 \epsilon}}{(-s)^\epsilon}} + 
	\sqrt { \frac{ \lambda \mu^{2 \epsilon}}{(-t)^\epsilon}} - \frac{ \epsilon^2 \sqrt{\lambda}}{8} \ln^2 \br{\frac{s}{t}}}
}

\section{Conclusions}
Thus, we found that the difference between the two areas reduces to a factor, which is equal to
\be
	\frac{Area_\sigma}{Area_{NG}} = \frac{2}{|\sin{\phi}|} \br{1 + \frac{\epsilon^2}{8}}
\ee
2 in the numerator is due to $S_\sigma$ instead of $\frac{1}{2} S_\sigma$, which is actually related to $S_{NG}$, was used. For orthogonal $\ok_a$, and for Alday-Maldacena's solution particularly $\sin{\phi}$ = 1, and substitution of one action by another in the scattering amplitude results only in constant phase shift, which is equal to $\frac{\sqrt{\lambda}}{8 \pi}$. So, physics of the process is left unaffected by this substitution. This gives us a hope, that in further investigations and applications one action can be replaced with the other one. However, it still remains unknown whether this simple relation between amplitudes would survive if one takes into account further terms of the semiclassical expansion.

\section*{Acknowledgements}
We would like to thank A.Mironov and A.Morozov for discussions. This work is partly supported by Federal Agency of Atomic Energy of Russia, by grant for support of scientific schools NSh-8004.2006.2 and by grant RFBR 07-02-00878.

\end{document}